\documentclass[twocolumn,preprintnumbers,aps,showpacs,superscriptaddress,prl]{revtex4-1}

\usepackage{palatino}
\usepackage[latin1]{inputenc}
\usepackage{epsf}
\usepackage{amsmath,amssymb}
\usepackage{latexsym}
\usepackage{calc}
\usepackage{color}
\usepackage{shadow}
\usepackage{epsfig}
\usepackage{float}

\usepackage[pdftex, pdfstartview = {FitH}]{hyperref}

\newcommand{\ben}{\begin{equation}}
\newcommand{\een}{\end{equation}}
\newcommand{\bea}{\begin{eqnarray}}
\newcommand{\eea}{\end{eqnarray}}

\def\dulR{{\underline{\underline{\bf R}}}}
\def\dulr{{\underline{\underline{\bf r}}}}

\begin{document}
  \title{Steps in the exact time-dependent potential energy surface} 
  \author{Ali Abedi}
\affiliation{Max-Planck Institut f\"ur Mikrostrukturphysik, Weinberg 2,
D-06120 Halle, Germany}
\author{Federica Agostini}
\affiliation{Max-Planck Institut f\"ur Mikrostrukturphysik, Weinberg 2,
D-06120 Halle, Germany}
\author{Yasumitsu Suzuki} 
\affiliation{Max-Planck Institut f\"ur Mikrostrukturphysik, Weinberg 2,
D-06120 Halle, Germany}
\author{E.K.U. Gross}
\affiliation{Max-Planck Institut f\"ur Mikrostrukturphysik, Weinberg 2,
D-06120 Halle, Germany}

\date{\today}
\begin{abstract}
We study the exact Time-Dependent Potential Energy Surface (TDPES) in the presence
of strong non-adiabatic coupling between the electronic and nuclear motion. The concept of the TDPES emerges from the exact factorization of the full
electron-nuclear wave-function [A. Abedi, N. T. Maitra, and E. K. U. Gross, Phys. Rev. Lett. \textbf{105}, 123002 (2010)]. Employing a 1D
model-system, we show that the TDPES exhibits a dynamical step that bridges between piecewise adiabatic shapes. We analytically investigate the position of the steps 
and the nature of the switching between the adiabatic pieces of the TDPES.
\end{abstract}

\maketitle 
The description of coupled electron-nuclear motion is one of the biggest
challenges in condensed-matter physics and theoretical chemistry. 
Fundamental to our understanding is the adiabatic separation of electronic and nuclear
motion embodied in the Born-Oppenheimer (BO) approximation. It allows one 
to visualize -approximately- a molecule as a set of nuclei moving on a single
Potential Energy Surface (PES) generated by the electrons in a specific
electronic eigenstate. The BO approximation breaks down when two or more
BOPES come close or cross. Some of the most fascinating and most challenging molecular
processes occur in the regime where the BO approximation is not valid,
e.g. ultrafast nuclear motion through conical intersections~\cite{domcke-yarkony},
radiationless relaxation of excited electronic states~\cite{wohlgemuth,*sobolewski}, intra- and
inter-molecular electron and proton transfer~\cite{rose,*martinez,*hanna}, to name a few. The
standard way of studying and interpreting these, so-called,
"non-adiabatic" processes is to expand the full molecular wave function
in terms of the BO electronic states. Within this expansion, non-adiabatic
processes can be viewed as a nuclear wave packet with contributions on
several BOPESs, coupled through the non-adiabatic coupling (NAC) terms 
which in turn induce transitions between the BOPESs. While this provides a 
formally exact description one may nevertheless ask: Is it also possible to study 
the molecular process using a {\it single} PES? This question is particularly
relevant if one thinks of a classical or semi-classical treatment of the
nuclei where a well-defined single classical force would be highly desirable.

In a recent Letter, we have introduced an exact time-dependent potential
energy surface (TDPES) that, together with an exact time-dependent vector
potential govern the nuclear motion. These concepts emerge from a novel
way to approach the coupled electron-nuclear dynamics via an exact 
factorization of the electron-nuclear wave function~\cite{AMG}. Features of the exact TDPES were
studied in the presence of strong laser fields~\cite{AMG,AMG2}. In the present Letter we
investigate the generic features of the exact TDPES {\it without} external 
laser but in the presence of strong non-adiabatic couplings. A major result will
be that the exact TDPES exhibits nearly discontinuous steps connecting
different static BOPES, reminiscent of Tully's surface hopping [6] in the
classical limit.

In~\cite{AMG} we have proved that the {\it exact} solution of the time-dependent Schr\"odinger equation (TDSE),
\ben\label{eq:tdse}
 \hat{H} \Psi(\dulr,\dulR,t)= i\partial_t\Psi(\dulr,\dulR,t),
\een
of the complete system of interacting electrons and nuclei can be written as a single product (unlike the BO expansion), 
$\Psi(\dulr,\dulR,t)=\Phi_{\dulR}(\dulr,t)\chi(\dulR,t)$, of the nuclear wave-function, $\chi(\dulR,t)$, and the electronic conditional
wave-function, $\Phi_{\dulR}(\dulr,t)$, that satisfies the partial normalization condition (PNC), 
$\int d\dulr\vert\Phi_{\dulR}(\dulr,t)\vert^2=1$. In the absence of time-dependent external fields, the system is described by the
Hamiltonian $\hat{H}$,
\begin{equation}
 \hat{H} = \hat{H}_{BO}(\dulr,\dulR) +\hat{T}_n(\dulR),
\end{equation}
that contains the traditional BO electronic Hamiltonian, $\hat{H}_{BO}(\dulr,\dulR) = \hat{T}_e(\dulr) + \hat{W}_{ee}(\dulr) + 
\hat{V}_{en}(\dulr,\dulR) + \hat{W}_{nn}(\dulR)$, and the nuclear kinetic energy, $\hat{T}_n(\dulR)$. Throughout this paper we use atomic
units (unless stated otherwise) and the electronic and nuclear coordinates are collectively denoted by $\dulr$ and $\dulR$, respectively.

The exact electronic wave-function satisfies the equation        
\ben\label{eq:exact_el_td}
 \Bigl(\hat{H}_{BO}(\dulr,\dulR)+\hat U_{en}^{coup} -\epsilon(\dulR,t)\Bigr)\Phi_{\dulR}(\dulr,t)\\=i\partial_t \Phi_{\dulR}(\dulr,t),
\een
where the electron-nuclear coupling operator is $\hat U_{en}^{coup}=\sum_{\nu=1}^{N_n}\big((-i\nabla_\nu-{\bf
A}_\nu(\dulR,t))^2/2+\left(-i\nabla_\nu \chi/\chi+{\bf A}_\nu(\dulR,t)\right)\left(-i\nabla_\nu-{\bf
A}_\nu(\dulR,t)\right)\big)/M_\nu$. The time-evolution of the nuclear wave-function is governed by the Schr\"odinger equation:
\ben\label{eq:exact_n_td}
 \Bigl(\sum_{\nu=1}^{N_n}\frac{(-i\nabla_\nu+{\bf A}_\nu(\dulR,t))^2}{2M_\nu} +\epsilon(\dulR,t)\Bigr)\chi(\dulR,t)=i\partial_t
 \chi(\dulR,t).
\een
These equations lead to rigorous definitions of the TDPES and the time-dependent vector potential
\bea
 \label{eq:exact_eps_td}
 &&\epsilon(\dulR,t) = \epsilon_{gi}(\dulR,t) + \epsilon_{gd}(\dulR,t) \\
 \label{eq:exact_vect}
 &&{\bf A}_\nu(\dulR,t)=\left\langle\Phi_{\dulR}(t)\right\vert\left.-i\nabla_\nu\Phi_\dulR(t)\right\rangle_\dulr.
\eea
The TDPES consists of two parts: $\epsilon_{gi}(\dulR,t)$, defined as
\ben\label{eq:exact_eps_gi}
 \epsilon_{gi}(\dulR,t) = \left\langle\Phi_{\dulR}(t) \right\vert\hat{H}_{BO}(\dulr,\dulR)+\hat U_{en}^{coup}((\dulr,\dulR,t)\left\vert \Phi_{\dulR}(t)\right\rangle_\dulr,
\een
is form-invariant under the gauge-transformation 
$\Phi_{\dulR}(\dulr,t)\rightarrow\tilde{\Phi}_{\dulR}(\dulr,t)=\exp(i\theta(\dulR,t))\Phi_{\dulR}(\dulr,t)$, $\chi(\dulR,t)\rightarrow\tilde
{\chi}(\dulR,t)=\exp(-i\theta(\dulR,t))\chi(\dulR,t)$, whereas $\epsilon_{gd}(\dulR,t)$, defined as 
\ben\label{eq:exact_eps_gd}
 \epsilon_{gd}(\dulR,t) =  \left\langle\Phi_{\dulR}(t) \right\vert - i \partial_t\left\vert \Phi_{\dulR}(t)\right\rangle_\dulr,
\een
is the part that depends on the choice of the gauge. Here, $\langle ..|..|..\rangle_\dulr$ denotes an inner product over the  
electronic variables only.

Why is this representation of the correlated electron-nuclear many-body problem exciting? The wave-function $\chi(\dulR,t)$ that satisfies 
the exact nuclear equation of motion~(\ref{eq:exact_n_td}) leads to an $N$-body density $\Gamma(\dulR,t)=\vert\chi(\dulR,t)\vert^2$ and an 
$N$-body current density ${\bf J}_\nu(\dulR,t)=Im(\chi^*\nabla_\nu\chi)+\Gamma(\dulR,t){\bf A}_\nu$ which reproduce the true nuclear $N$-body density and current density obtained from 
the full wave-function $\Psi(\dulr,\dulR,t)$~\cite{AMG2}. In this sense, $\chi(\dulR,t)$, can be viewed as the proper nuclear wave-function. The 
time evolution of $\chi(\dulR,t)$, on the other hand, is completely determined by the TDPES, $\epsilon(\dulR,t)$, and the vector potential, ${\bf A}_\nu(\dulR,t)$. 
Moreover, these potentials are {\it unique} up to within a gauge transformation. This uniqueness is straightforwardly proven by following the steps of the 
current-density version~\cite{Ghosh-Dhara} of the Runge-Gross theorem~\cite{RGT}. In other words, if one wants a TDSE~(\ref{eq:exact_n_td}) whose solution 
$\chi(\dulR,t)$ yields the true nuclear $N$-body density and current density, then the potentials appearing in this TDSE are (up to within a gauge 
transformation) uniquely given by Eqs.~(\ref{eq:exact_eps_td}-\ref{eq:exact_eps_gd}); there is no other choice. This also implies, that the gradient of 
this exact TDPES {\it is the only correct force on the nuclei} in the classical limit (plus terms arising from the vector potential, if those cannot be 
gauged away). The goal of this Letter is to find out how this exact TDPES looks like when one has strong non-adiabatic couplings in the traditional 
expansion in BO states. One major result will be that the exact TDPES shows a nearly discontinuous step whenever the nuclear wave-packet splits in 
the vicinity of an avoided crossing of the BOPES.  
\begin{figure}[h]
 \centering
 \includegraphics*[width=1.0\columnwidth]{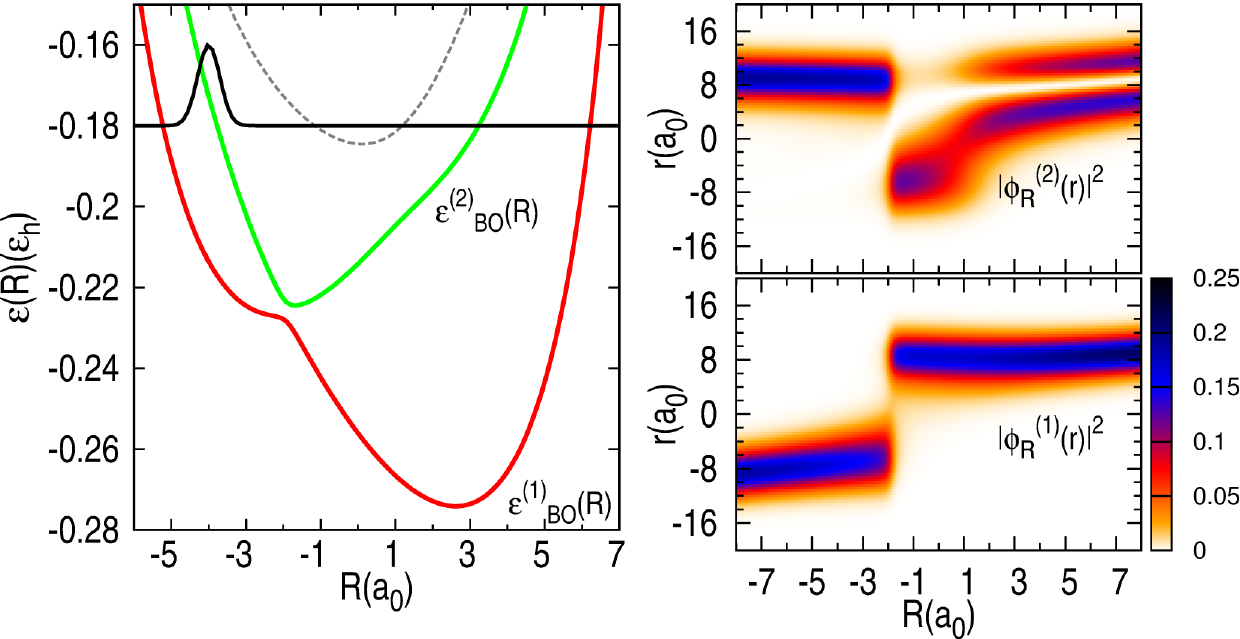}
 \caption{Left: The first two BOPESs (indicated in the figure) together with the 3rd BOPES (black dashed-line) and the initial nuclear
  wave-function (black solid-line) . Right: Adiabatic electronic conditional densities as indicated in the figures}
 \label{fig:BO-data}
\end{figure}

To study the exact TDPES we first of all need a problem that is simple enough to allow for a numerically exact solution and that nevertheless 
exhibits the characteristic features associated with strong non-adiabatic couplings, such as the splitting of the nuclear wave packet. For this 
purpose we employ the model of Shin and Metiu~\cite{MM}. It consists of three ions and a single electron. Two ions are fixed at a distance of
$L=19.0~a_0$, the third ion and the electron are free to move in one dimension along the line joining the two fixed ions. The Hamiltonian
of this system reads
\bea
 \label{eq:metiu-hamiltonian}
 \begin{split}
  &\hat{H}(r,R) =  - \frac{1}{2}\frac{\partial^2}{\partial r^2}-\frac{1}{2M}\frac{\partial^2}{\partial R^2}  + \frac{1}{\vert \frac{L}{2}
  - R \vert } + \frac{1}{\vert \frac{L}{2} + R \vert} \\
  &- \frac{\mathrm{erf}\left(\frac{\vert R-r \vert}{R_f}\right)}{\vert R - r \vert} - \frac{\mathrm{erf}\left(\frac{\vert r-\frac{L}{2}
  \vert}{R_r}\right)}{\vert r-\frac{L}{2}  \vert} -
  \frac{\mathrm{erf}\left(\frac{\vert r+\frac{L}{2} \vert}{R_l}\right)}{\vert r+\frac{L}{2} \vert}.
 \end{split}
\eea
Here, the symbols $\dulr$ and $\dulR$ are replaced by $r$ and $R$, the coordinates of the electron and the movable nucleus measured from the center of the two fixed ions. 
$M=1836$~a.u. and we choose $R_f=5.0~a_0$, $R_l=3.1~a_0$ and $R_r=4.0~a_0$
such that the first BOPES, $\epsilon^{(1)}_{BO}$, is strongly coupled to the second BOPES, $\epsilon^{(2)}_{BO}$, around the 
avoided crossing at $R_{ac}=-1.90~a_0$ and there is a weak coupling to the rest of the surfaces. The first three BOPES are shown in
Fig.~\ref{fig:BO-data} (left panel), together with the BO conditional electronic densities $\vert \phi^{(1)}_R(r)\vert^2$  and $\vert
\phi^{(2)}_R(r)\vert^2$ (right panels). As expected, $\vert \phi^{(1)}_R(r)\vert^2$  and $\vert \phi^{(2)}_R(r)\vert^2$
exhibit abrupt changes, along the $R$-axis, at the position of the avoided crossing, $R_{ac}$: $\vert \phi^{(1)}_R(r)\vert^2$
switches from being localized around the fixed ion on the left ($r=-9.5~a_0$), to be localized around the one on the right ($r=9.5~a_0$); $\vert \phi^{(2)}_R(r)\vert^2$ on the other hand, presents a single-peak structure for $R<R_{ac}$ and a
double-peak structure for $R>R_{ac}$.

\begin{figure}[h]
 \centering
 \includegraphics*[width=1.0\columnwidth]{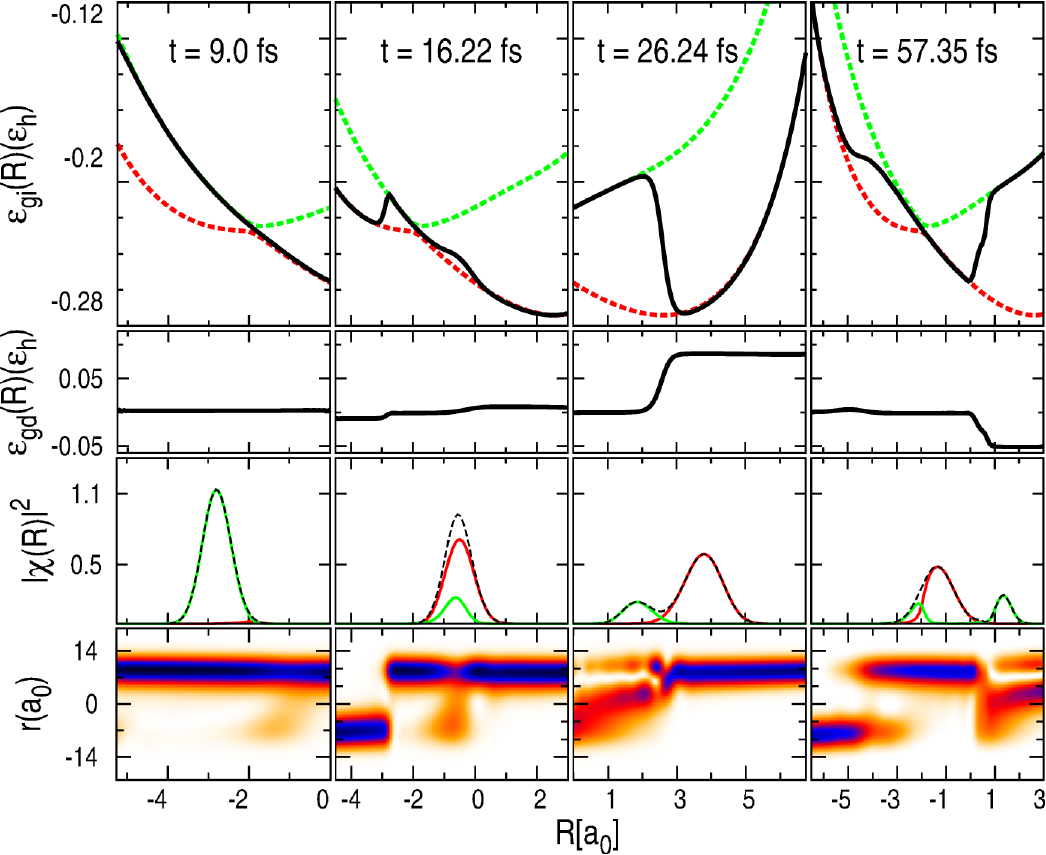}
 \caption{First panel (top): The gauge independent part of the TDPES  (black solid-line)
plotted at four different times (indicated), $\epsilon^{(1)}_{BO}$ (red
dashed-line) and $\epsilon^{(2)}_{BO}$ (green dashed-line). Second panel
(from the top): the gauge dependent part of the TDPES is plotted at the same times.
Third panel (from the top): the exact nuclear density (black dashed-line)
is shown together with $\vert F_1(R,t) \vert^2$ (red solid-line) and
$\vert F_2(R,t) \vert^2 $ (green solid-line).
Lowest panel: the exact time-dependent electronic conditional density,
$\vert \Phi_R(r,t) \vert^2$, is plotted. The color range is the same as
Fig.~\ref{fig:BO-data}.}
  \label{fig:snapshots}
\end{figure}

We suppose that the system is initially excited to $\epsilon^{(2)}_{BO}$ and the initial nuclear wave-function is a
wave-packet with the width $\sigma=1/\sqrt{2.85}$, centered at $R=-4.0~a_0$ (see Fig.~\ref{fig:BO-data}, black solid-line), i.e., the
initial full wave-function is $\Psi_0(r,R) =Ae^{-(R-4)^2/\sigma^2}\phi^{(2)}_R(r)$ with $A$ being a normalization constant.
Starting with $\Psi_0(r,R)$ as initial state, we propagate the TDSE, numerically exactly, to obtain the full molecular wave-function
$\Psi(r,R,t)$ and from it we calculate, as discussed in Ref.~\cite{AMG2}, the TDPES in the gauge where the vector potential is zero. 
Hence, the TDPES is the only potential acting on the nuclear sub-system. In the upper panel of
Fig.~\ref{fig:snapshots}, the gauge-invariant part of the TDPES~(\ref{eq:exact_eps_gi}), $\epsilon_{gi}$, is plotted (black
solid-line) at four different times, along with the two lowest BOPESs, $\epsilon^{(1)}_{BO}$ (red dashed-line) and $\epsilon^{(2)}_{BO}$ (green
dashed-line). In the second panel (from the top), the gauge-dependent part of the TDPES~(\ref{eq:exact_eps_gd}), $\epsilon_{gd}$, is
plotted at the same times. In the third panel (from the top), the exact nuclear density (black dashed-line), $\vert \chi(R,t)
\vert^2$, is shown together with the absolute value squared of the projection of the full wave-function on the first and second BO
electronic states, i.e., $\vert F_1(R,t) \vert^2=\vert\int dr \phi^{(1)*}_R (r)\Psi(r,R,t)\vert^2$ (red solid-line) and $\vert F_2(R,t)
\vert^2 =\vert \int dr \phi^{(2)*}_R(r) \Psi(r,R,t)\vert^2$ (green solid-line). In the lowest panel, $\vert \Phi_R(r,t) \vert^2$ is presented.

At the initial time ($t=0$), due to the choice of the initial state, the TDPES coincides with $\epsilon^{(2)}_{BO}$. Since $\Psi_0(r,R)$ is not an eigenstate of the
Hamiltonian~(\ref{eq:metiu-hamiltonian}), it evolves in time. At $t=9.0 fs$, $\epsilon_{gi}$ coincides with $\epsilon^{(2)}_{BO}$ for
$R<R_{ac}$, goes smoothly through the avoided crossing region and follows $\epsilon^{(1)}_{BO}$ for $R>R_{ac}$, resembling the {\it diabatic} PES 
of state 2 in Ref.~\cite{MM}, in which the electron interacts with the fixed ion on the right ($r=9.5~a_0$) and
with the moving ion, but not with the fixed ion on the left ($r=-9.5~a_0$). As $\epsilon_{gd}$ is constant in this region
(Fig.~\ref{fig:snapshots}), the TDPES is identical with $\epsilon_{gi}$ \footnote{In Fig.~\ref{fig:snapshots}, curves representing $\epsilon_{gd}$ have been 
rigidly shifted along the energy axis}. The nuclear wave-packet is driven by the
TDPES to spread towards the avoided crossing of two BOPESs, where a significant non-adiabatic transition happens and the exact nuclear
density splits. Already at this moment, a slight transition of the nuclear wave-packet to the lower surface is visible around the avoided
crossing. At later times, e.g., $t=16.22~fs$, $t=26.24~fs$ and $t=57.35~fs$., far from the avoided crossing, $\epsilon_{gi}$ contains 
steps that connect its different pieces that are on top of different BOPESs in different slices of $R$-space.
In the region around $R_{ac}$, it follows the diabatic surface that passes smoothly through the avoided crossing. On the 
other hand, $\epsilon_{gd}$ is piecewise constant and presents similar steps as $\epsilon_{gi}$. Therefore, the TDPES, $\epsilon_{gi}+\epsilon_{gd}$, preserves 
the features mentioned before, i.e., (i) far from the avoided crossing, it presents steps 
that connect the regions in $R$-space in which the TDPES has the
shape of one BOPES to the regions in which it has the shape of the other BOPES; (ii) around the avoided crossing, it
follows the diabatic surface that smoothly connects one BOPES to the other.

The exact TDPES represented in Fig.~\ref{fig:snapshots} can be viewed from a different perspective. The nuclear wave-packet from a
semi-classical point of view can be represented as an ensemble of classical trajectories, along which point-particles evolve under the
action of a classical force which is the gradient of $\epsilon_{gi}$. According to our observations, on different sides of a step such a
force is calculated from different BOPESs. This is reminiscent of the {\it jumping between the
adiabatic surfaces} in algorithms such as Tully's surface hopping~\cite{TSH, SCND1}. However, while Tully surface hopping is a stochastic algorithm, 
the jumps in the exact TDPES correspond to an exact solution of the TDSE. When the time-dependent vector potential can not be set to zero, a gauge can be chosen in which $\epsilon_{gd}$ is zero and a 
time-dependent vector potential together with $\epsilon_{gi}$ specifies the classical force that the nuclei experience in different slices of
$R$-space. Investigating $\epsilon_{gi}$ together with the time-dependent vector potential for a wide range of situations may help to improve the existing 
semi-classical procedures to simulate non-adiabatic nuclear dynamics.

The exact time-dependent electronic conditional density, shown in the lower panels of Fig.~\ref{fig:snapshots} at different times, behaves
similarly to the TDPES: (i) it smoothly connects a $\vert\phi^{(2)}_R(r)\vert^2$-like structure, by crossing $R_{ac}$, with a
$\vert\phi^{(1)}_R(r)\vert^2$-like structure, or vice versa, presenting a diabatic behavior, e.g. at $t=9.0~fs$; (ii) it displays abrupt changes, between 
regions that piecewise match different adiabatic conditional densities.

\begin{figure}[hb!]
 \centering
 \includegraphics*[width=0.80\columnwidth]{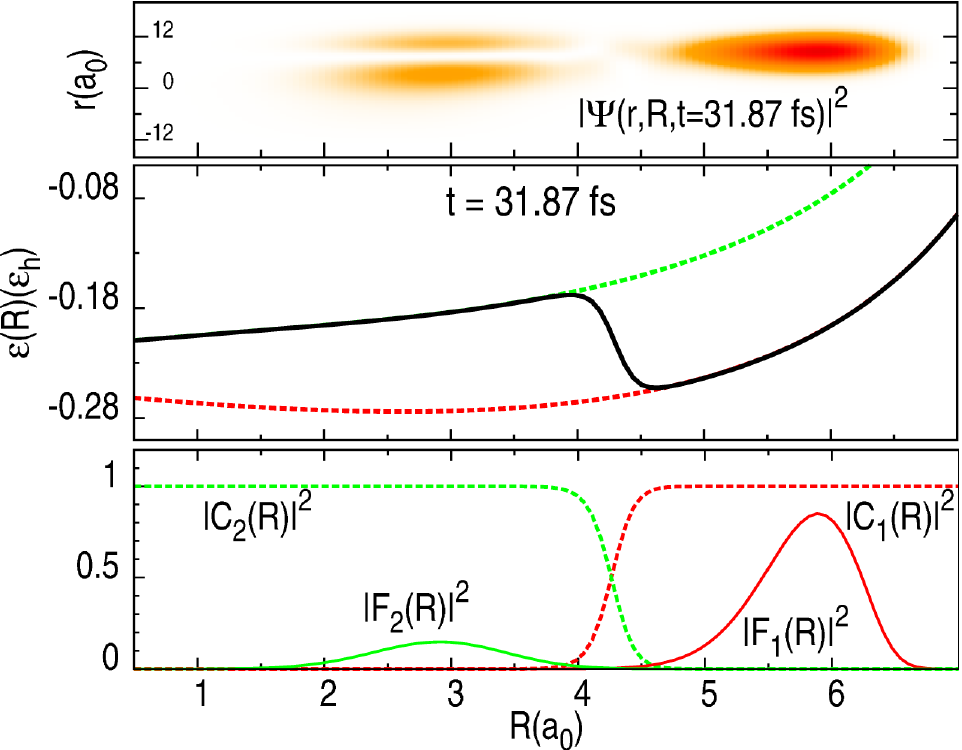}
\caption{Top: the full electron-nuclear density at the $t=31.87~fs$.
Middle: a snapshot of the gauge invariant part of the TDPES (solid black line) at the
$t=31.87~fs$. For reference, $\epsilon^{(1)}_{BO}$ (red dashed-line) and
$\epsilon^{(2)}_{BO}$ (green dashed-line) are shown.
   Bottom: Expansion coefficients (indicated in the figure) of the (two
states) adiabatic expansion of the full wave-function and the exact
electronic conditional wave-function (see the text) at the $t=31.87 fs$.}
 \label{fig:details-1600}
\end{figure}

In order to analyze the behavior of the TDPES, we rewrite it by expanding the exact electronic
conditional wave-function in terms of the adiabatic electronic states ~\cite{AMG2}. Due to the choice of the parameter in the Hamiltonian,
we only need to include the first two BO states, then 
\ben\label{eqn: electronic wf expansion}
 \Phi_R(r,t) = C_1(R,t)\phi_R^{(1)}(r) + C_2(R,t)\phi_R^{(2)}(r).
\een
We expand the full electron-nuclear wave-function in the same basis,
\ben\label{eqn: total wf expansion}
 \Psi(r,R,t) = F_1(R,t)\phi_R^{(1)}(r) + F_2(R,t)\phi_R^{(2)}(r),
\een
where the expansion coefficients, $F_k$'s and $C_k$'s, are related as
\begin{equation}\label{eqn:coffs}
 C_k(R,t) = \frac{F_k(R,t)}{\chi(R,t)} = 
 \frac{e^{-i\theta(R,t)}F_k(R,t)}{\sqrt{\left|F_1(R,t)\right|^2+\left|F_2(R,t)\right|^2}}.
\end{equation}
Here, $\theta$ is the phase of the exact nuclear wave-function and we have used the relation $\vert\chi(R,t)\vert^2=\vert
F_1(R,t)\vert^2+\vert F_2(R,t)\vert^2$, determined by $\int dr \vert \Psi(r,R)\vert^2$ using Eq.~(\ref{eqn: total 
wf expansion}) and the orthonormality of the adiabatic states. By using Eqs.~(\ref{eq:exact_eps_td}) and (\ref{eqn: electronic wf 
expansion}), we rewrite $\epsilon_{gi}(R,t)$ and $\epsilon_{gd}(R,t)$ in terms of $\epsilon_{BO}^{(k)}(R)$ and  $C_k(R,t)$ ($k=1,2$)
\begin{eqnarray} 
 \epsilon_{gi}(R,t)&=&\sum_{k=1,2}\vert C_k(R,t)\vert^2\epsilon_{BO}^{(k)}(R)\label{eqn:BO-expansion 1}\\
 \epsilon_{gd}(R,t)&=&\sum_{k=1,2}\vert C_k(R,t)\vert^2\dot{\gamma}_k(R,t),\label{eqn:BO-expansion 2}
\end{eqnarray}
where $\gamma_1$ and $\gamma_2$ are the phases of $C_1,\,C_2$. In Eq.~(\ref{eqn:BO-expansion 1}), all terms of $\mathcal O(M^{-1})$ have been
neglected and it only contains BOPESs which are the leading terms responsible for the shape of $\epsilon_{gi}(R,t)$, especially far from
the avoided crossing where the NACs are small. The gauge-dependent term is written in terms of the time derivative of the phases,
$\dot{\gamma_1}$ and $\dot{\gamma_2}$. $\vert C_1 \vert^2$ and $\vert
C_2 \vert^2$ vary between 0 and 1 and $\vert C_1 \vert^2 + \vert C_2 \vert^2 =1$ by virtue of the PNC. Therefore, as
Eq.~(\ref{eqn:BO-expansion 1}) suggests, in the region where $\epsilon_{gi}(R,t)$ coincides with $\epsilon_{BO}^{(1)}(R)$, the 
corresponding expansion coefficient $\vert C_1 \vert^2$ is close to one while $\vert C_2 \vert^2$ is close zero and vice versa. We have
observed (Eq.~(\ref{eqn:coffs})) that at $R_0$, the cross-over of $\vert F_1 \vert$ and $\vert F_2 \vert$ where
$\left|F_1(R_0,t)\right|=\left|F_2(R_0,t)\right|=\left|X(t)\right|$, $\left|C_1\right|^2$ and 
$\left|C_2\right|^2$ are always equal to $1/2$ and $R_0$ is the center of the region where steps
form. Moving away from this point, one of the $\left|C_k\right|^2$'s becomes dominant (Fig.~\ref{fig:details-1600}) and $\epsilon_{gi}(R,t)$
lies on top of the corresponding BOPES.

To elaborate on how the TDPES switches between the two adiabatic states, we Taylor expand $\vert C_k(R,t) \vert^2$ around $R_0$ and keep
only up to the linear order terms 
\begin{equation}\label{eqn: linear F}
 \left|C_{\mathop{}_2^1}(R,t)\right|^2=\left[1 \pm \alpha(t)\left(R-R_0\right)\right]/2,
\end{equation}
where
\ben
 \alpha(t)=\frac{\left(\nabla_R\left|F_1(R,t)\right|\right)_{R_0}-\left(\nabla_R\left|F_2(R,t)\right|\right)_{R_0}}{\left|X(t)\right|}.
\een
Eq.~(\ref{eqn: linear F}), using the relation $0 \leq \vert C_k \vert^2 \leq 1$ ($k=1,2$) or equivalently $\left|R-R_0\right|\leq\alpha^{-1}$, estimates 
the width of the region, $\Delta R$, where the switching between BOPESs occurs with $\frac{2}{\alpha}$, i.e., $\Delta R = \frac{2}{\alpha}$. Hence, the 
larger the values of $\alpha$, the sharper the steps become.

As an example, we discuss the TDPES at $t=31.87~fs$ in Fig.~\ref{fig:details-1600}. As it is seen, $\epsilon_{gi}$
switches from  $\epsilon_{BO}^{(1)}(R)$ to  $\epsilon_{BO}^{(2)}(R)$ over the region where $\vert F_1\vert$ and $\vert F_2\vert$ cross
(see the bottom plot). As $\vert F_1 \vert$ and $\vert F_2 \vert$ have opposite slopes and cross where they are small, $\alpha$ is
large yielding a small $\Delta R$. Outside the switching region, one of the $|C_k|^2$s becomes dominant. Interestingly, 
the exact electron-nuclear density contains signatures of the behavior $\epsilon_{gi}$, i.e., where $\epsilon_{gi}$ coincides 
with $\epsilon_{BO}^{(1)}(R)$, presents one peak in analogy with $\vert \phi^{(1)}_R(r)\vert^2$ (see Fig.~\ref{fig:BO-data}), while, 
where $\epsilon_{gi}(R,t)$ follows $\epsilon_{BO}^{(2)}(R)$, it displays two peaks, like $\vert\phi^{(2)}_R(r)\vert^2$ (Fig.~\ref{fig:BO-data})
The step of $\epsilon_{gi}$ in the intermediate region is indicated by scars in the full electron-nuclear density.

In conclusion, we have presented generic features of the exact TDPES for situations in which, according to the standard BO expansion
framework, significant non-adiabatic transitions occur and the nuclear wave-packet splits at the avoided crossing of two BOPESs. For the
$1$D model system studied here,  the TDPES is the only
potential that governs the dynamics of the nuclear wave-function (the vector potential can be gauged away) and provides us with an alternative 
way of visualizing and interpreting the non-adiabatic processes. 
We have shown that the gauge-invariant part of the TDPES, $\epsilon_{gi}(R,t)$, is characterized by two generic features: (i) in the vicinity 
of the avoided crossing, $\epsilon_{gi}(R,t)$, becomes identical with a diabatic PES in the direction of the wave-packet motion, (ii) far from the avoided crossing, $\epsilon_{gi}(R,t)$, as
a function of $R$, is piecewise identical with different BOPESs and exhibits nearly discontinuous steps in between. The latter feature holds after the wave-packet branches and leaves the avoided crossing. The gauge-dependent part, $\epsilon_{gd}(R,t)$, on the other hand, is piecewise constant in the region where $\epsilon_{gi}(R,t)$ coincides with different BOPESs. Hence $\epsilon_{gd}(R,t)$ has little effect on the gradient of the total TDPES, but may shift the BOPES-pieces of $\epsilon_{gi}(R,t)$ by different constants causing the exact TDPES to be piecewise parallel to the BOPESs. These features of the TDPES support the use of diabatic surfaces as the driving potential when 
a wave-packet approaches a region of strong NAC. Moreover, they are in agreement with the semi-classical picture of non-adiabatic nuclear dynamics that 
suggests to calculate the classical forces acting on the nuclei according to {\it the gradient of only one of the BOPESs}. We expect that these findings will ultimately lead to improved algorithms for the mixed quantum-classical treatment of electrons and nuclei.

This study was supported by the European Commission within the FP7 CRONOS project (ID 280879).

\bibliography{./steps}

\end{document}